\documentclass[multphys,vecphys]{svmult}

\usepackage{makeidx}     % allows index generation
\usepackage{graphicx}    % standard LaTeX graphics tool
\usepackage{multicol}    % used for the two-column index
\makeindex             % used for the subject index
\begin{document}

\title*{$^{56}$Ni mass in type IIP SNe: Light curves and H$\alpha$ luminosities diagnostics}
\titlerunning{$^{56}$Ni mass in type IIP SNe} 
\author{A. Elmhamdi\inst{1} \and  N.N. Chugai \inst{2} \and I.J. Danziger\inst{3}}
%\institute{Name and Address of your Institute
%\texttt{name@email.address}
%\and Name and Address of your Institute \texttt{name@email.address}}
\institute{SISSA / ISAS , via Beirut 4 - 34014 Trieste - Italy \texttt{elmhamdi@sissa.it}
      \and Institute of Astronomy RAS,
  Pyatnitskaya 48, 109017 Moscow - Russia
	\and Osservatorio Astronomico di Trieste, Via G.B.Tiepolo 11 - I-34131
	 Trieste - Italy }

\maketitle
\abstract{
 We analyse late-time observations, available photometry and spectra, of a sample of type II plateau supernovae (SNe IIP). The possibility of using H$\alpha$ luminosity at the nebular epoch as a tracer of $^{56}$Ni mass in this class of objects is investigated, yielding a consistency with the photometry-based estimates within 20\%. Interesting correlations are found and their impacts on our present understanding of the physics of core collapse SNe are discussed.    
}
\section{Results and discussion}
%\label{sec:1}
% Always give a unique label
% and use \ref{<label>} for cross-references
% and \cite{<label>} for bibliographic references
% use \sectionmark{}
% to alter or adjust the section heading in the running head
%Your text goes here. Use the \LaTeX\ automatism for your citations
%\cite{monograph}.
%
The study of SNe~IIP (i.e. optical properties, asymmetry, clumping, nucleosynthesis and yields) provides constraints on the 
explosion models and pre-supernova parameters. In particular, the $^{56}$Ni mass is one of the crucial parameters since it 
presumably depends on the presupernova structure and the 
explosion model \cite{Auf91}.

We select a sample of type IIP SNe on the basis of available photometry and spectra, especially at latter epochs. In Fig. 1(left) we display
 the absolute $V-$light curves of the SNe sample
 together with that of SN 1987A. We adopt unique distance determination methodology, namely using the recession velocity of the host galaxy  
 corrected for Local Group infall onto the Virgo Cluster and assuming a Hubble constant $\rm H_{0}$=70 km s$^{-1}$Mpc$^{-1}$. Galactic extinction is removed using the map
of galactic dust extinction by Schlegel et al (1998)\cite{Sch98}, while the host galaxy reddening is estimated from the ``$B-V$'' and ``$V-R$'' colour excess 
compared to the intrinsic colour curves of SN 1987A. 
This is based on the fact that at the late photospheric 
phase, through the end of the recombination phase, SNe IIP seem to
follow colour evolution similar to SN 1987A \cite{Schm92}.

The computed late time decline slopes, in the $150-400$ d time range, are consistent with the radioactive decay of $^{56}$Co and consequent trapping 
 of the gamma-rays. Indeed a mean value of about 
 $<\gamma^{V} >~ \simeq ~0.99~ (\pm 0.13)$ for the sample SNe is measured. 
%%%%%%%%%%%%%%%%%%%%%%%%%%%%%%%%%%%%%
\begin{figure}
\centering
\includegraphics[height=6cm,width=6cm]{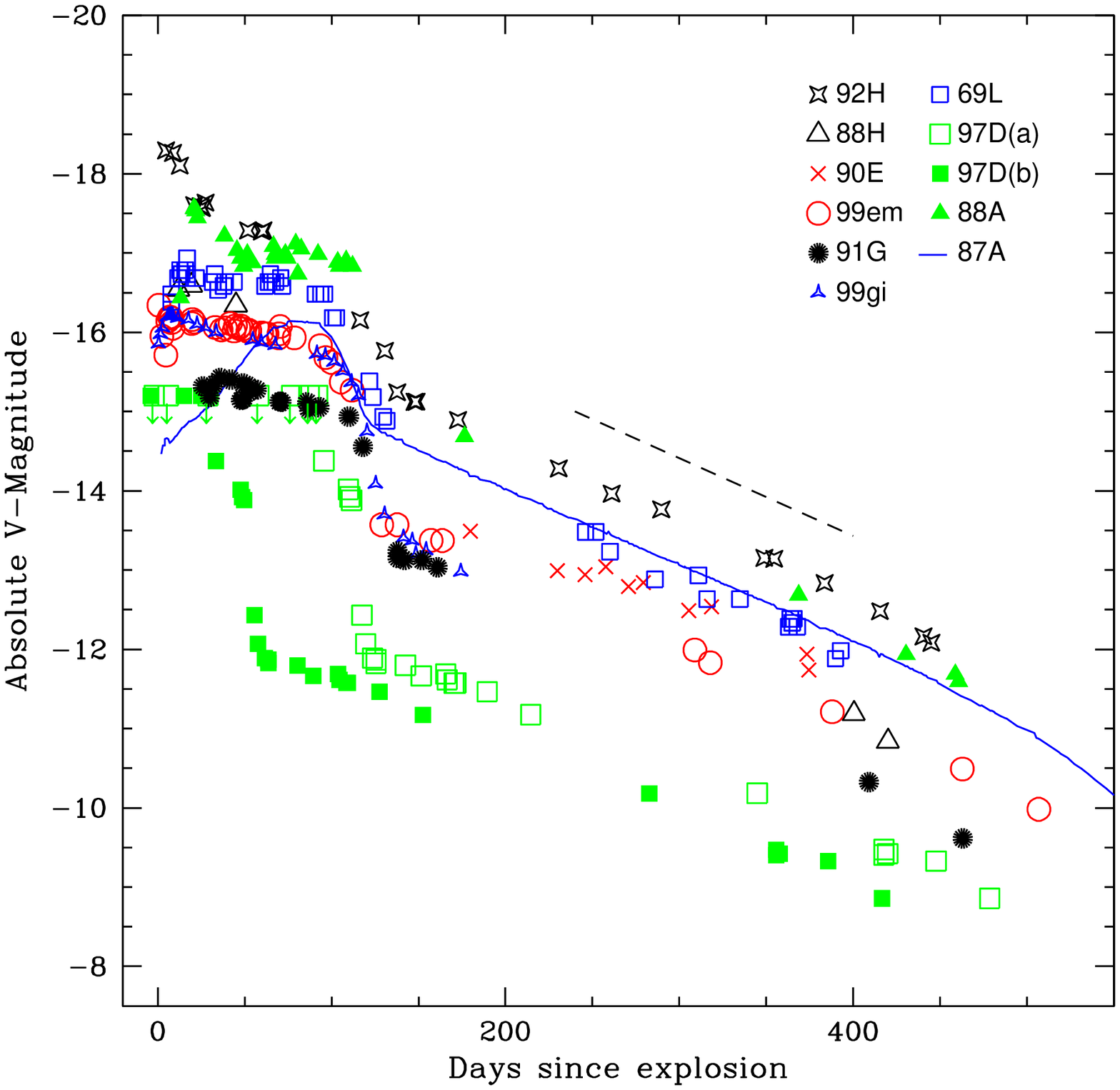}\includegraphics[height=6cm,width=6cm]{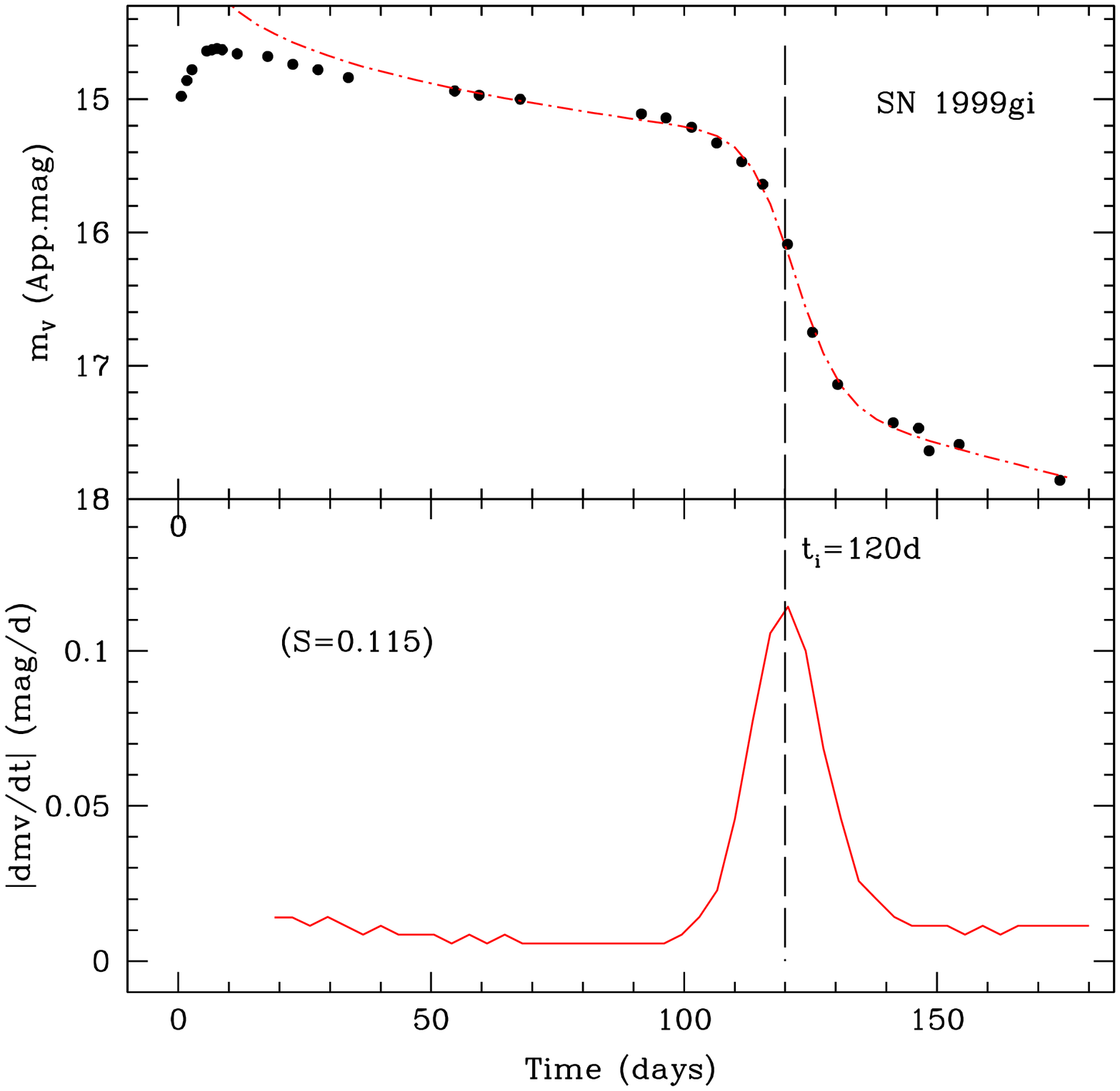}
\caption{Left: $M_V$ light curves of the SNe sample. The dashed line shows the slope for $^{56}$Co decay. Right: determination of the steepness and inflection time for the case of SN 1999gi (data from Leonard et al. 2002\cite{Leo02}).}
\label{fig:1}       % Give a unique label
\end{figure}
%%%%%%%%%%%%%%%%%%%%%%%%%%%%%%%%%%%%%

Once we fix the points related to the extinction and distance, which are crucial when dealing with a SN-sample study, we proceed with computing the amounts of ejected $^{56}$Ni using the absolute $V-$light curve of SN 1987A as template (in the 120$-$400 days time range). We found a range from lower values for SN 1999eu and SN 1997D to a higher one for SN 1992H ($[0.003-0.13]~M_\odot$), with an average of about 0.05 $M_\odot$. These results combined with  the emergence of the extremely faint and bright events (see also Pastorello et al.; Hamuy., these Proceedings) tend to change the general belief about the $^{56}$Ni production in type II SNe (i.e. all eject $\sim 0.07~ M_\odot$). The above facts are extremely important for modeling the chemical evolution of galaxies since they impose constraints on the iron production.

Analysing the absolute light curves of the sample events, we introduce a new parameter, called ``$steepness$'' and defined as $S=-dM_V/dt$ , which describes the shape of the light curves and provides a way to measure the decay rate at the inflection point. The light curves in the transition period from plateau to 
radioactive tail are approximated by a sum of plateau and radioactive terms:

\begin{equation}
 F= A\frac {(t/t_{0})^{p}}{1 + (t/t_{0})^{q}}+B\exp (-t/111.26),    
\end{equation}

\noindent where $A$, $B$, $t_{0}$, $p$ and $q$ are parameters derived by 
the $\chi ^{2}$ minimization technique in the sensitive 
interval $t_{\rm i}\pm50$ days.
Fig. 1(right) demonstrates the behaviour of $S$ and determination of the inflection time $t_i$ for the case of SN 1999gi. 

The correlation between the $^{56}$Ni mass and plateau M$_V$ found by Hamuy \cite{Ham03} is confirmed (Fig. 2, left), and the linear fit is described by the equation:

\begin{equation}
\log\,M(^{56}{\rm Ni})=-0.438M_V(t_{\rm i}-35)-8.46. 
\end{equation}

Furthermore an interesting by-product of the sample photometry analysis is the correlation found between $^{56}$Ni mass and the steepness parameter $S$. The correlation is quantified on the basis of the available data (Fig. 2, right), and the best linear fit reads:
\begin{equation}
\log\,M( ^{56}{\rm Ni}) = -6.2295\,S -0.8147
\end{equation}

The correlation is such that the steeper the decline at the inflection point the lower is the mass of $^{56}$Ni. Although the interpretation of this correlation requires hydrodynamical modeling with different amounts of
$^{56}$Ni and degrees of mixing, it may well be that somehow the increase
of the $^{56}$Ni mass in SNe~IIP ejecta favours the larger radiative diffusion times at the end of the plateau and, therefore, a
less steep transition from plateau to the radioactive tail, or 
 that the increase of the $^{56}$Ni mass is accompanied by the growth
 of the degree of mixing which favours a less steep decline. 
This correlation is interesting in the sense that, if confirmed, it will provide distance and extinction independent estimates of the $^{56}$Ni mass in SNe IIP.
%%%%%%%%%%%%%%%%%%%%%%%%%%%%%%%%%%%%%
\begin{figure}
\centering
\includegraphics[height=6cm,width=6cm]{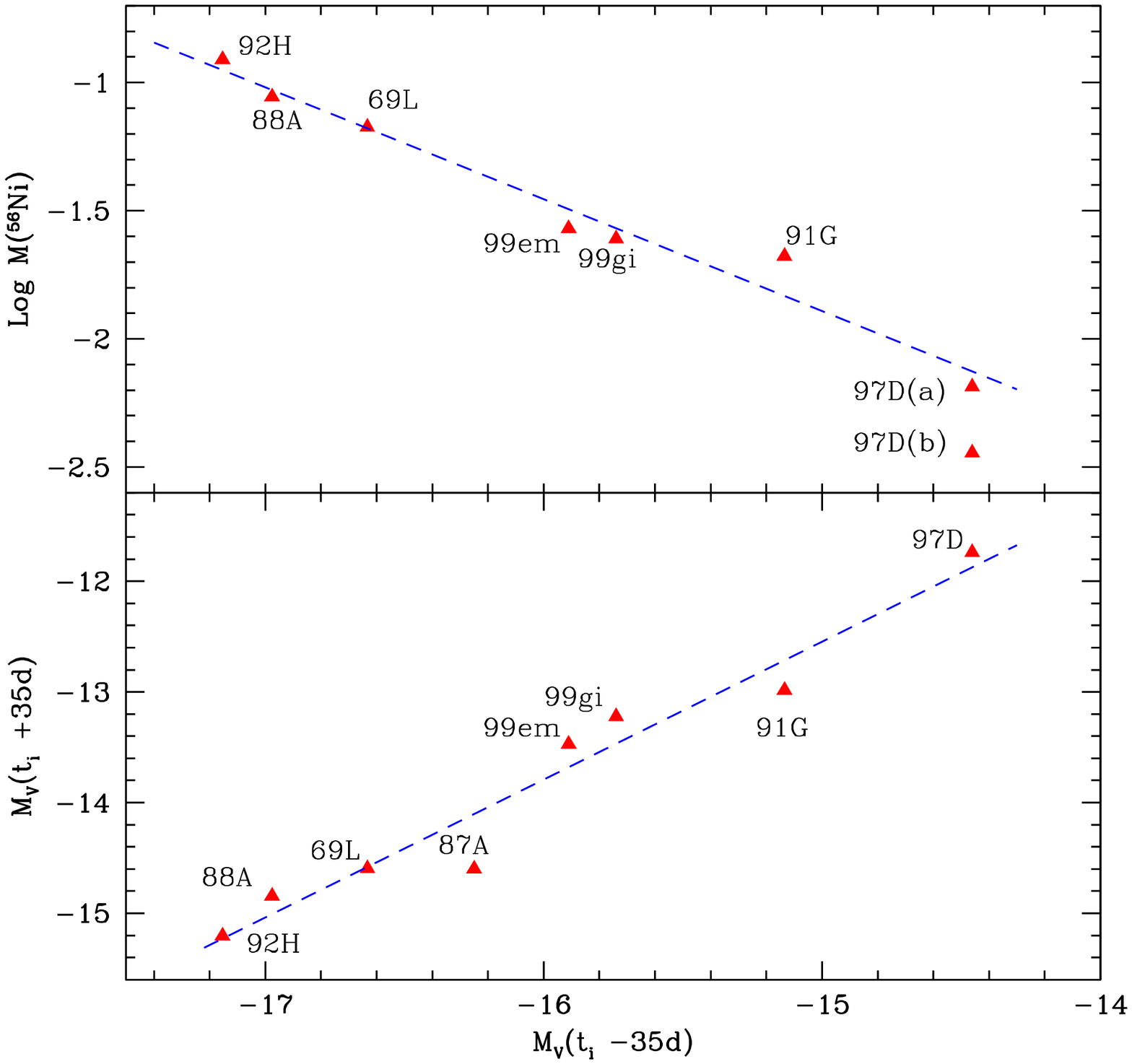}\includegraphics[height=6cm,width=6cm]{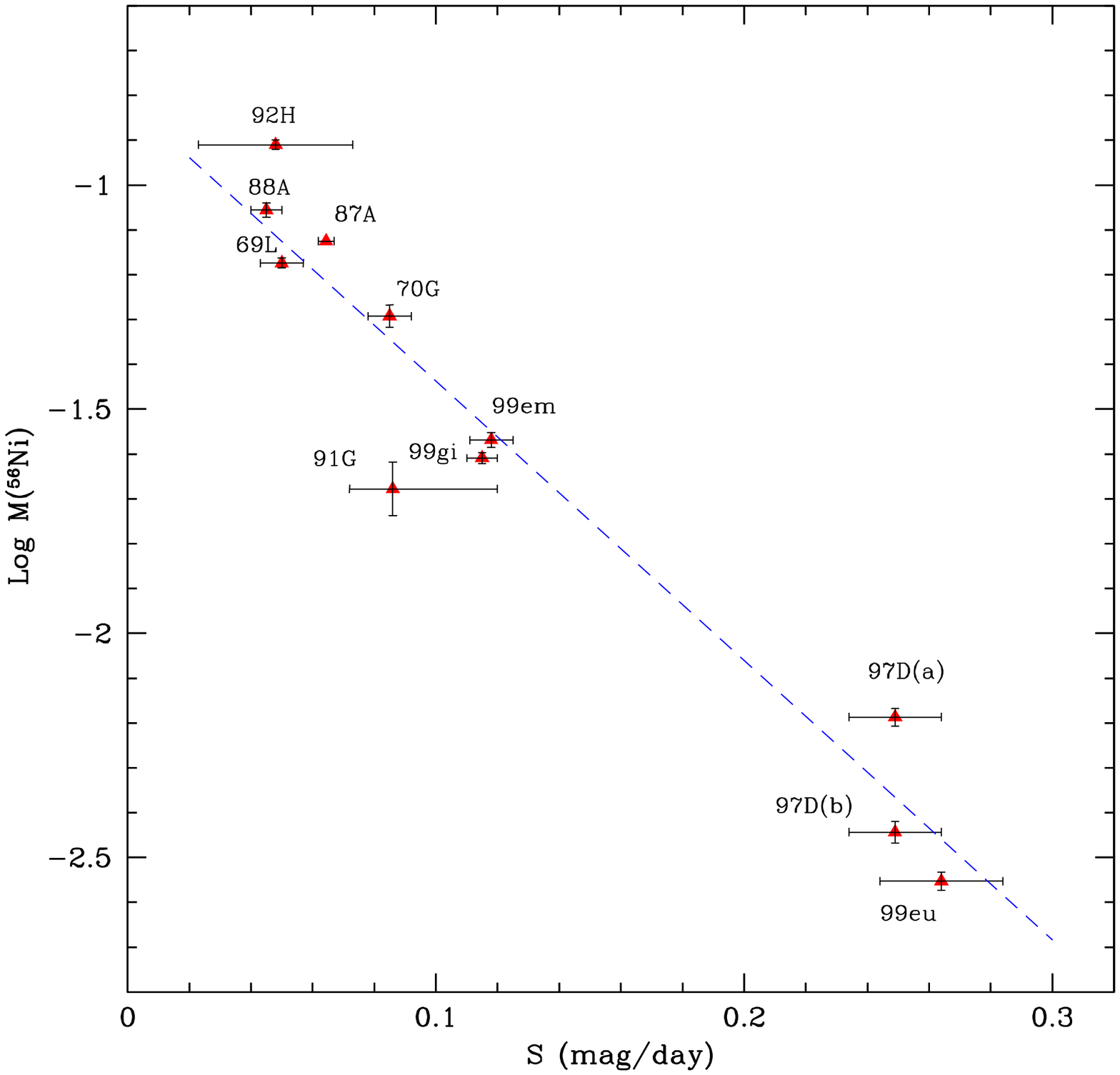}
\caption{Left: the correlation between $M_V$ magnitudes of plateau and 
of radioactive tail. Upper panel shows the correlation between 
 $M_V$ at the moment ($t_{\rm i}-35$d) and $^{56}$Ni mass derived 
 from the tail magnitudes. Lower panel shows directly the 
 correlation of plateau magnitude $M_V(t_{\rm i}-35{\rm d})$ and tail 
 magnitude $M_V(t_{\rm i}+35{\rm d})$. Right: the correlation between $^{56}$Ni mass and steepness $S$. Case ``a'' of SN 1997D is adopted for the fits (dashed lines). }
\label{fig:2}       % Give a unique label
\end{figure}
%%%%%%%%%%%%%%%%%%%%%%%%%%%%%%%%%%%%%
             
We construct then a ``$two-zone$'' model of the H$\alpha$ luminosity in 
SN~IIP to explore the sensitivity of the H$\alpha$ behaviour to 
variation of model parameters. The primary purpose of the upgraded model is to specify better the early nebular phase compared to the previous version \cite{Chug90}. We found that if mass, energy and 
mixing conditions do not vary strongly among SNe~IIP  
(less than factor 1.4) then with an accuracy better than 10\% 
H$\alpha$ luminosity is proportional to $^{56}$Ni mass during 
the $200-400$ days after explosion (for more details see Elmhamdi et al. 2003 \cite{Elm03}).
H$\alpha$ luminosities are then used to derive $^{56}$Ni masses of the SNe sample. This is done employing two approaches: first, 
using the H$\alpha$ light curve in SN~1987A as template and, second, 
applying the model fitting (Fig. 3). Both approaches agree  
within 15\% unless we are dealing with extreme cases such as SN~1970G
(type IIP/L) and the underluminous SN~1997D.  
In both these cases we should possess additional information 
about ejecta mass and energy to derive the $^{56}$Ni mass from H$\alpha$ modeling. SN 1997D is indeed a special case for which two scenarios have been argued, namely: a small age scenario with a low ejecta mass (case ``b''; \cite{Chug00}) and a large age option with high mass of the ejecta (case ``a''; \cite{Zamp03}). 
%%%%%%%%%%%%%%%%%%%%%%%%%%%%%%%%%%%%%
\begin{figure}
\centering
\includegraphics[height=7cm,width=8cm]{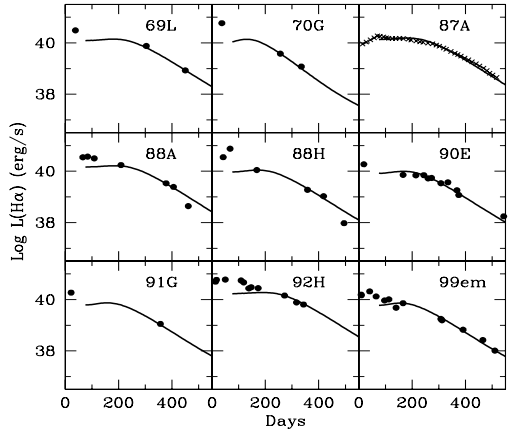}
\caption{The H$\alpha$ luminosity evolution in different SN~IIP.
  Models ({\em solid lines}) are overplotted on the observational 
  data {\em dots}.}
\label{fig:3}       % Give a unique label
\end{figure}
%%%%%%%%%%%%%%%%%%%%%%%%%%%%%%%%%%%%%

Worth noting is the simple approach of using H$\alpha$ light curve of SN 1987A as a template to estimate the  $^{56}$Ni mass. 
The $^{56}$Ni mass values derived using this method agree within 20\% with those from the photometry, which thus gives us 
confidence that H$\alpha$ is a good indicator of the amount of 
$^{56}$Ni in SNe~IIP (Fig. 4). Simultaneously, this consistency suggests that parameters of SNe~IIP (mass, energy and mixing) are not very 
different. In fact this is consistent with the uniformity of 
plateau luminosities and plateau lengths of SNe~IIP. 

This simple approach is applied then for three SNe for which we have late spectra but no photometry (SNe 1995ad, 1995V and 1995W), giving reasonable values, and thus demonstrating the usefulness of the method \cite{Elm03}.
Generally, the approach based upon H$\alpha$ 
 may be indispensable in cases, when 
the photometry at the nebular epoch is absent, or when there is 
a problem with subtraction of stellar background (SN~IIP in the bulge, 
or in high redshift galaxies).
%%%%%%%%%%%%%%%%%%%%%%%%%%%%%%%%%%%%%
\begin{figure}
\centering
\includegraphics[height=7cm,width=7cm]{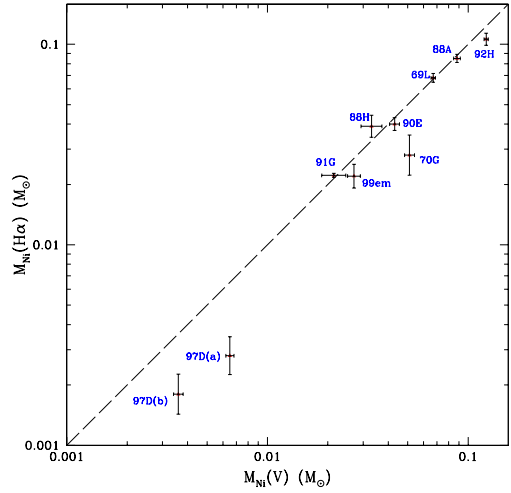}
\caption{The correlation between the $^{56}$Ni mass derived from the H$\alpha$ luminosity, $M_{\rm Ni}(\mbox{H}\alpha)$, and that 
 derived from the tail $M_V$ magnitude, $M_ {\rm Ni}(V)$.
The dashed line has a slope of unity. Clear deviation is seen for SN 1970G(SNP/L) and for the faint event SN 1997D. }
\label{fig:4}       % Give a unique label
\end{figure}
%%%%%%%%%%%%%%%%%%%%%%%%%%%%%%%%%%%%%

On the one hand, the clustering of
the points around two values of $^{56}$Ni mass viz. 0.005 and 0.05 $\rm M_{\odot}$ in Fig. 4 may result from poor statistical sampling, or it may be a hint that a mechanism such as fall-back is an important one in the evolution of the low-mass group.

The interesting correlations demonstrated in this class of objects point to a high degree of homogeneity and this is encouraging for the use of type IIP SNe as cosmological probes. Improved statistical samples and better sampled data are however needed to have a firmer confirmation of these correlations.
   
These kinds of analyses demonstrate how large sample analyses in SNe studies can provide robust results and correlations with their consequent impact for our understanding of SNe physics.

%%%%%%%%%%%%%%%%%%%%%%%% referenc.tex %%%%%%%%%%%%%%%%%%%%%%%%%%%%%%
% sample references
% "physics"
%
% Use this file as a template for your own input.
%
%%%%%%%%%%%%%%%%%%%%%%%% Springer-Verlag %%%%%%%%%%%%%%%%%%%%%%%%%%

%
% BibTeX users please use
% \bibliographystyle{}
% \bibliography{}

%%%%%%%%%%%%%%%%%%%%%%%%%%%%%%%%%%%%%%%%%%%%%%%%%%%%%%%%%%%%%%%%%%%%%%  }

%%%%%%%%%%%%%%%%%%%%%%%%%%%%%%%%%%%%%%%%%%%%%%%%%%%%%%%%%%%%%%%%%%%%%%

\printindex
\end{document}